\documentclass[10pt]{article}

\usepackage{makeidx}  
\usepackage[dvips]{graphicx}
\usepackage{dsfont}
\usepackage{amsmath}
\usepackage{algorithm}
\usepackage{algorithmic}

\usepackage{cite}
\usepackage{url}  
\usepackage{epsfig}
\usepackage{color}
\usepackage{psfrag}
\usepackage{amssymb}

\usepackage[english]{babel}

\usepackage{rotating}
\usepackage{rotate}
\usepackage{lscape}

\usepackage{multirow}

 \usepackage{graphicx}
\usepackage{dcolumn}
\usepackage{bm}

\def\lim{\mathop{\rm lim}}

\newcommand{\mypar}{\par\vspace{0,1cm}\noindent}

\newtheorem{definition}{Definition}[section]

\DeclareSymbolFont{bbold}{U}{bbold}{m}{n}
\DeclareSymbolFontAlphabet{\mathbbold}{bbold}

\makeatletter
\newcommand*{\rom}[1]{\expandafter\@slowromancap\romannumeral #1@}

\usepackage[labelfont=bf,labelsep=period,justification=raggedright]{caption}

\makeatletter
\renewcommand{\@biblabel}[1]{\quad#1.}
\makeatother

\date{}

\begin{document}

\begin{flushleft}
{\Large
\textbf{Quantitative Graph Theory: A new branch of graph theory and network science}
}
\\
Matthias Dehmer$^{1,2 \ast}$, Frank Emmert-Streib$^{3,4 \ast}$, and Yongtang Shi $^{5,6}$ 
\\
\bf{} $^1$Department of Computer Science, Universit\"at der Bundeswehr M\"unchen, Germany \\
\bf{} $^2$Department of Mechatronics and Biomedical Computer Science,
UMIT, Hall in Tyrol, Austria \\
\bf{} $^3$Computational Medicine and Statistical Learning Laboratory, Department of Signal Processing, Tampere University of Technology, Finland \\
\bf{} $^4$Institute of Biosciences and Medical Technology, 33520 Tampere, Finland\\
\bf $^5$Center for Combinatorics and LPMC-TJKLC, Nankai University, Tianjin 300071, China\\
\bf $^6$ College of Computer and Control Engineering, Nankai University, Tianjin 300071,
China\\

\end{flushleft}

\begin{abstract}

In this paper, we describe {\sc quantitative graph theory} and argue it is a new graph-theoretical branch in network science, however, with significant different features compared to classical graph theory.
The main goal of quantitative graph theory is the structural quantification of information contained in complex networks by employing a {\it measurement approach} based on numerical invariants and comparisons. Furthermore, the methods as well as the networks do not need to be deterministic but can be statistic. As such this complements the field of classical graph theory, which is descriptive and deterministic in nature. We provide examples of how quantitative graph theory can be used for novel applications in the context of the overarching concept network science. 

\end{abstract}

\section{Introduction}\label{sec_intro}
Graph theory \cite{harary} is a relatively new branch of mathematics. Examples of major activities include investigating topological aspects of graphs by, e.g., Kuratowski \cite{kura} and Gross \cite{gross_1987}, leading to the
emergence of topological graph theory \cite{gross_1987}, studying embeddings of graphs in surfaces and graphs as topological spaces. A highlight
in this theory is surely the theorem of Kuratowski for characterizing planar graphs. Further activities contributed numerous methods for exploring structural properties of networks, e.g., see \cite{bollabas,dehmer_birk_book_2010,halin,harary}. Note that there are also other
branches of Graph Theory, such as Extremal or Random Graph Theory, which have also been very fruitful \cite{bollabas}.

Based on those multifaceted methods, numerous problems by means of graphs
have been solved in disciplines such as artificial intelligence and pattern
recognition \cite{Conte_2004}, biology \cite{emmert_chronique_fatigue,junker_2008},
chemoinformatics \cite{randic_1979,randic_1990,SubMat2_varmuza_2000},
cognitive modeling \cite{sommerfeld_1990}, computational
linguistics \cite{Mehler:Weiss:Luecking:2010:a},
image recognition \cite{Hsieh_2008,theoharatos_2004}, machine learning \cite{borgwardt_2007,horvath1} and
web mining \cite{dehmer_tatra}.

In a wider sense, graph-based approaches have been used extensively in various disciplines. The hype dealing with complex networks also contributed a lot to modern Graph or Network Theory
and has been triggered from the breakthrough
of the world wide web and other physically-oriented studies exploring networks (graphs) as complex systems \cite{albert_1999,WaSt98}.
Besides investigating only random graph models for analyzing complex systems, it turned out that there is a strong need to further develop {\textit{quantitative}} approaches.
A main reason for this was the insight that many real-world networks are composed of non-random topologies where quantitative
methods such as graph measures \cite{Costa_2007,dehmer_birk_book_2010} have been proven essential to quantify structural information of graphs.
\mypar
When studying the existing literature dealing with classical aspects of graph theory \cite{halin,harary},
it turns out that most of the existing contributions are {\textit{descriptive}} approaches for describing graphs structurally.
Examples thereof are Kuratowski's theorem mentioned above, the description of Eulerian paths, graph colorings and so forth \cite{behzad,halin,harary}.
This leads us straightforwardly to a definition of {\textit{Quantitative Graph Theory}}:
\begin{definition}
Quantitative graph theory (QGT) deals with the quantification of structural aspects of graphs, instead of characterizing graphs only descriptively,
\end{definition}
We see that the aspect of {\textit{measurement}} is crucial here. To quantify structural information of a graph means
to employ any kind of measurement, i.e., a local or global one.
\mypar
Concrete examples for quantitative approaches in the context of Graph Theory are, for instance:
\begin{itemize}

\item Graph similarity or distance measures \cite{dehmer_tatra,sobik1,zelinka1}.

\item Graph measures to characterize the topology of a graph~\cite{Costa_2007,dehmer_birk_book_2010,todeschini_2002}. This group of measurements includes numerical graph invariants
which are also referred to as
{\textit{topological indices}} \cite{bonchev_book_1991,DeBa00}.

\item The exploration of metrical properties of graphs and the derivation of measures  \cite{SkDo88}.

\item Consideration of structural graph measures as network complexity measures and exploring properties thereof \cite{bonchev_2009,dehmer_mowshowitz_emmert_1_2011}.

\item Information Theory and Statistics of graphs \cite{arnand_2009,kolaczyk_2009}.

\item Applications of the above graph measures in social network analysis, chemistry, biology and other disciplines \cite{DeBa00,wasserman}.

\item Using machine learning for deriving quantitative approaches for graphs \cite{cook_2007}.

\end{itemize}

\mypar
As known, various graph measures have been defined and investigated \cite{Costa_2007,dehmer_birk_book_2010}
even in the early fifties \cite{bavelas}. Seminal contributions from this development have been graph centrality measures for investigating problems in sociology, e.g.,
group performance \cite{bavelas}. Numerous other graph-theoretical measurements have been developed afterwards, but to date, there have been no attempts to consider Quantitative Graph Theory as a graph-theoretical branch.
In view of the vast number of existing descriptive methods  \cite{behzad,halin,harary}, quantitative techniques to analyze graphs are clearly underrepresented so far.
However, quantitative approaches for exploring graphs have been examined from different perspectives in a variety of disciplines including Discrete Mathematics, Computer Science,
Biology, Chemistry and related areas \cite{dehmer_emmert_book_2009,junker_2008}.
Interestingly, in Biology, Computer Science and Discrete Mathematics, the focus has tended to be on the comparative analysis of graphs such as graph similarity and graph distances. In contrast, 
in Chemistry and related fields, the principal aim has been to quantify structural features of graphs leading to numerical graph invariants \cite{diudea}.

\begin{figure}[t!]
\center
\includegraphics*[scale=0.8]{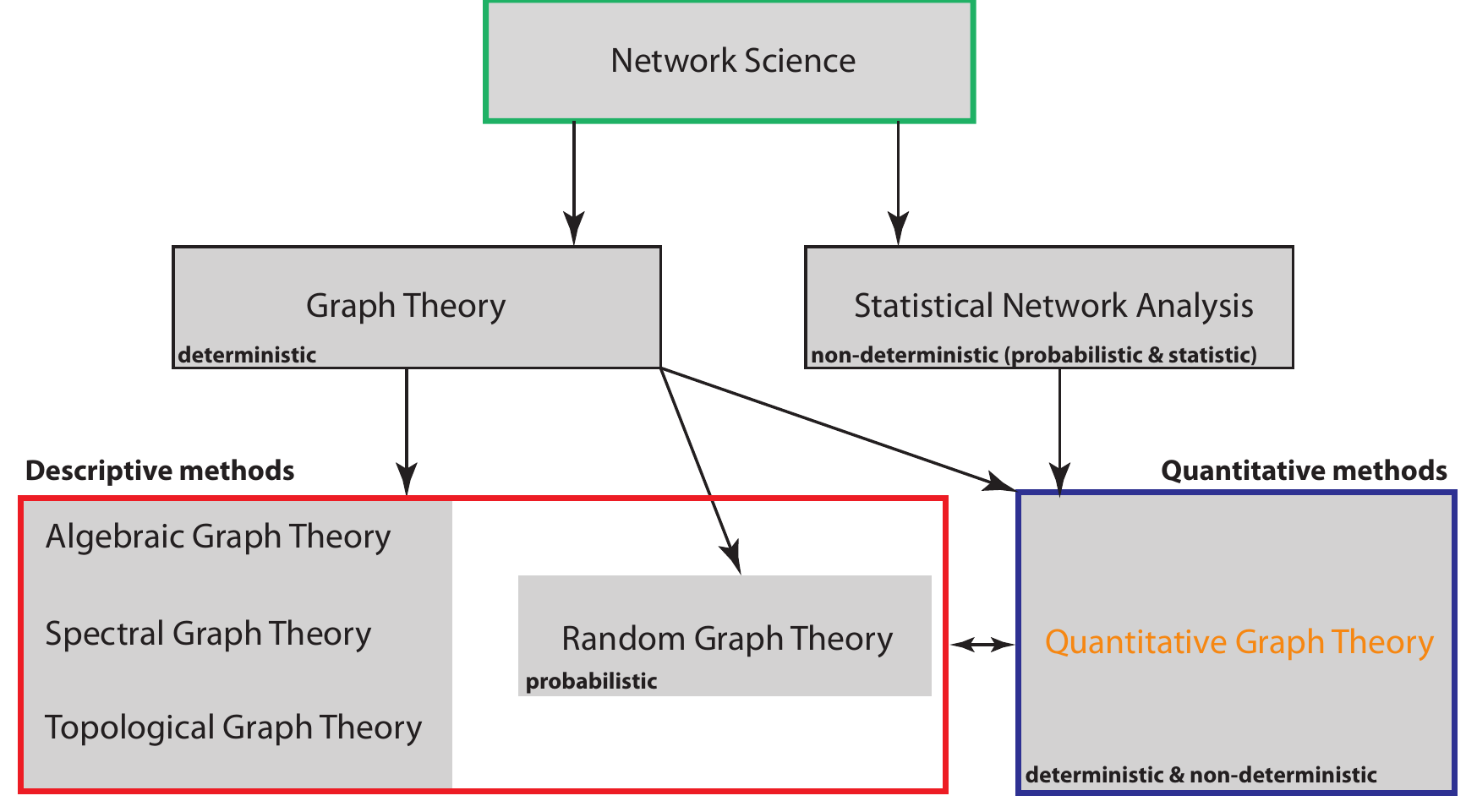}
\caption{Quantitative graph theory as a branch of graph theory and network science.}\label{pic_1}
\end{figure}

\section{Quantitative Graph Theory}
In \cite{dehmer_CRC_book_chapter_2014}, Quantitative Graph Theory has been defined as a measurement approach to quantify structural information
of networks. In general, local, global or comparative graph measures can be employed for measuring structural information. See also Section \ref{sec:methods_quant}. 
We emphasize that this definition complements 
(classical) graph theory which mainly deals with the description of structural properties of graphs, see \cite{harary}. Examples for pure descriptive graph theory methods are
graph colorings, graph embeddings and decompositions, and characterizations of graphs like the Theorem of Kuratowski \cite{kura}.
In contrast, pure quantitative methods are quantitative graph measures
that are based on estimating structural features of networks \cite{dehmer_birk_book_2010}. The latter class of methods can be interpreted as graph complexity measures.
Seminal work in this area was done by Mowshowitz \cite{mowshowitz_1} and Bonchev \cite{bonchev_2009}. Note that the ambiguity of some methods belonging 
to both Descriptive and Quantitative Graph Theory has been explained in \cite{dehmer_CRC_book_chapter_2014}. 

In Figure \ref{pic_1}, we present a conceptual overview that shows the connection of Quantitative Graph Theory within some other related fields. First, we note that Network Science comprises both, Graph Theory and Statistical Network Analysis because it is the most generic field. The major difference between Graph Theory and Statistical Network Analysis is that the graphs as well as the methods of Graph Theory  
are purely deterministic, whereas in statistical network analysis the networks and the methods may be non-deterministic. A cause for the structural non-deterministicness may be given by, e.g., measurement errors or signal variability used to infer such networks \cite{emmert_chronique_fatigue}. Hence, the methods 
applied are based on probabilistic and statistical principles. Furthermore, Quantitative Graph Theory is also a branch of Graph Theory which relates to quantify structural
information of networks. This is in contrast to classical Graph Theory which mainly deals with descriptive graph analysis, as discussed above. Interestingly, one can see that
random graph theory is on one hand deterministic, when describing properties of random graphs. On the other hand, the methods and the random graphs themselves are often
probabilistic and, hence, this subfield is also non-deterministic in nature.

\section{Methods Used in Quantitative Graph Theory}\label{sec:methods_quant}

According to Dehmer et al. \cite {dehmer_CRC_book_chapter_2014}, Quantitative graph theory can be divided into two major categories, namely, comparative graph analysis \cite {dehmer_CRC_book_chapter_2014} and graph characterization by using numerical graph invariants is another category which relates to determine the structural complexity of networks. In the following, we give some prominent examples of methods for these two categories. For more details see \cite {dehmer_CRC_book_chapter_2014}.

\subsection{Comparative Graph Analysis}\label{sec:comparative}

Comparative graph analysis means determining the structural similarity or distance between two or more networks.
A problem has been choosing the right concept for this task meaningfully as a graph similarity measure depends on an underlying concept.
When doing so, two main classes thereof namely ({\textit{exact}} graph matching) \cite{zelinka1} 
and {\textit{inexact}} matching \cite{bunke4} to compare networks structurally have been developed. 
In particular, methods from exact graph matching have been demanding to use for large networks as the measures rely on graph isomorphism.
As known, the complexity  of the graph isomorphism has not yet determined for arbitraty networks. 

Comparative techniques for networks have been widely applied
in many scientific disciplines. Examples are artificial intelligence and pattern
recognition \cite{Conte_2004}, computational biology \cite{emmert_chronique_fatigue,junker_2008},
chemoinformatics \cite{randic_1990,SubMat2_varmuza_2000},
image recognition \cite{Hsieh_2008,theoharatos_2004}, and applied mathematics \cite{dehmer_tatra,dehmer_shi_2014}.  A subset thereof is:

\begin{itemize}
\item Isomorphism-based Measures \cite{kaden1,sobik1,zelinka1}.
\item Graph Edit Distance \cite{bunke1}
\item Iterative Graph Similarity Methods \cite{blondel_2004,zager_2008}
\item String-based Measures \cite{dehmer_tatra,emmert_streib1,robles_kelly} 
\item Graph Kernels \cite{borgwardt_2007,horvath1}
\end{itemize}

\subsubsection{Applications}
As many tasks in machine learning rely on similarity concepts, it has been straightforward to use graph similarity (comparative network analysis)
for applications. For instance, graph similarity measures can be used to compare and to classify structures representing real objects which can be visualized
as relational structures. When doing so, it has been also important to use graph similarty concepts in machine learning as the underlying algorithms
are often rely on similarity between objects to be processed. For instance, the following application areas turned out to be useful and important:

\begin{itemize}
\item Similarity of Document Structures \cite{butler1,dehmer_GT_MLMTA07}.
\item Tree Similarity \cite{jiang_1994,selkow_tree} 
\item Molecular Similarity \cite{maggiora_2004,randic_1979}.
\item Statistical Graph Matching \cite{shams_2001,theoharatos_2006}. 
\end{itemize}

It is obvious that for some specific problems, one needs special measures which are not require to operate on the entire network. An example for this are tree similarity measures \cite{jiang_1994,selkow_tree} 
that operate on trees only representing connected and acyclic networks. Using such special comparative measures may be beneficial to cluster document or RNA structures  that are represented by trees only. This gives us
an idea about the complexity of the network similarity problem as various graph classes exist when it comes to realize a particular application.

\subsection{Graph Characterization}\label{sec:invariants}

In quantitative graph theory \cite{dehmer_CRC_book_chapter_2014}, graph characterization relates to determine the complexity of a 
given network by using numerical graph invariants.
Graph invariants are graph measures to characterize graphs structurally which are invariant under isomorphism \cite{todeschini_2002}. 
Examples for local graph invarinats are vertex degrees or vertex eccentricities~\cite{SkDo88,harary}; global graph invariants 
are, for instance, graph entropy measures 
or measures which are based on distances in networks like the well-known Wiener
index \cite{todeschini_2002}. All those measures have in common that they are functions which map networks to the reals \cite{todeschini_2002}. 
A special kind of graph measures are so-called topological indices \cite{todeschini_2002}. Note that they have been used in 
mathematical and structural chemistry extensively to
characterize molecular structures based on their size, shape, branching and cyclicity  \cite{todeschini_2002}.  
Importantly, an undesired aspect of all structural graph measures is their degeneracy \cite{DeGrVa12,todeschini_2002};
that means, many of these graphs measures fail to discriminate the structure of non-isomorphic graphs by their values uniquely. 
This property often has a negative effect when determining the complexity
of networks numerically. Therefore it has been important to investigate the problem on a large scale to finally obtain a 
classification of graph measures according to their properties such as
degeneracy, structural interpretation and so forth, see \cite{DeGrVa12,todeschini_2002}. 
Based on graph invariants to be used, graph measures can be roughly divided into the following classes:

\begin{itemize}
\item Distance-based graph measures. Examples: Wiener index, eccentrcity \cite{Wi47,SkDo88}. 
\item Degree-based measures. Examples: Randi\'{c} index and Zagreb index \cite{Randic75,NiKoMiTr03}.
\item Eigenvalue-based measures. Examples: Graph energy and largest positive eigenvalue of trees \cite{LiShiGutman_book_y2012,lovasz_pelikan_1973}.
\item Information-theoretic measures. Examples: Vertex-orbit entropy due to Mowshowitz and partition-independent graph entropies \cite{DeMo10_history}.
 \end{itemize}
 
As to applications, these graph measures have been used in chemoinformatics (e.g., QSAR and QSPR) \cite{todeschini_2002}
for predicting biological or physico-chemical properties of molecular structures \cite{DeBa00,todeschini_2002}. 
Also, structural graph measures
have been employed to analyze phylogenetic properties of metabolic networks \cite{mazurie_2008}.
Moreover, graph measures got attention from areas outside chemistry and biology as structural systems representing networks
occur in many other disciplines too.
Examples are computational linguistics \cite{Mehler:2009:c:Langform}  and web structure mining \cite{chakrabarti2}
where web-based units have been explored by using structiral indices.  

For applications, it turned out to be important understanding the properties of graph measures properly. 
Otherwise, the úsage of those measure might misleading. We here summarize some important
properties as follows:

\begin{itemize}
\item Uniqueness of graph measures (i.e., degeneracy) \cite{DeGrVa12,todeschini_2002}.
\item Information inequalities describing interrelations of information-theoretic network measures \cite{DeMo10_inequalities}.
\item Structural interpretation \cite{dehmer_CRC_book_chapter_2014}.
\item Usefulness and quality of graph measures \cite{dehmer_CRC_book_chapter_2014,furtula_2013}.
\item Correlation of graph measures \cite{dehmer_CRC_book_chapter_2014}.
\end{itemize}

\subsection{Software for Quantitative Graph Analysis}
In this section, we briefly discuss free and comercial software to analyze complex networks.

\subsubsection{Programs developed for the Statistical programming Language \texttt{R}}
The multiparadigm language \texttt{R} \cite{R_software_2008} turned out to be useful for analyzing networks from various disciplines. 
An example is the \texttt{R}-package \texttt{NetBioV} \cite{tripathi_2014} for visualizing large biological networks. 
Other standard \texttt{R}-libraries for structural network analysis are \texttt{igraph} and \texttt{graph}. 
They contain many methods like the calculation of shortest paths in networks. However they only offer a very few
quantitative methods for characterizing networks. For this purpose, 
the \texttt{R}-package \texttt{QuACN} \cite{quacnPackage} has been developed which contains over 150 structural network measures.  
Further \texttt{R}-package for performing network analysis can be found in \cite{dehmer_CRC_book_chapter_2014,mueller2012book_dehmer_trajanoski}.

\subsubsection{Commercial Software}
A well-kown example of commercial software  for claculating structural and molecular indices (measures) is 
\texttt{Dragon} \cite{dragon_software_2004}. Actually it contains more than 4000 molecular descriptors \cite{todeschini_2002} for characterizing
molecular networks. This software has been proven useful for QSAR and QSPR \cite{DeBa00}. Other commercial pieces of software are, e.g.,  
the Sentinel Visualizer \cite{fmsasg.com_2015} and Sonamine \cite{sonamine_2009}. Sentinel Visualizer \cite{fmsasg.com_2015} is a software for analyzing
and visualizing social networks. Sonamine \cite{sonamine_2009} has been developed for performing data mining and simulation techniques 
for massive social networks.

\subsubsection{Conjecture engines}
Conjecture engines \cite{Graffiti,GrInvIn} are tools that have been used to generate conjectures when proving statements on topological graph measures.
They can be used to test the validity of inequalities involving topoological graph measures.
The tool \texttt{Graffiti}~\cite{Graffiti} is a prominent example and has been famous for generating conjectures involving the general and simple 
Randi\'{c} index. 
A succesor thereof is \texttt{GrInvIn}~\cite{GrInvIn} which is a freely available as \texttt{Java}-Application 
A shortcoming of \texttt{GrInvIn} is the quite weak conjecture engines which does not allow to use this tool to explore complex
conjectures.

\section{Summary and Conclusion}
In this chapter, we put the emphasis on quantitative graph theory. We pointed out that this intriguing field can be seen as a new branch of graph theory and should be treated as such. 
So far, most of the contributions in graph theory dealt with descriptive approaches for
describing structural features of networks, see Section \ref{sec_intro}. By demonstrating the potential of quantitative approaches which have been widely spread over
several disciplines, we believe that the discussed approaches will break new ground. An important finding of this chapter is that quantitative graph theory is per se
strongly interdisciplinary. The mathematical apparatus belongs to applied mathematics but, as demonstrated, 
quantitative graph theory have been applied in various scientific areas.

\section*{Acknowledgments}
Matthias Dehmer thanks the
Austrian Science Funds for supporting this work (project P26142).
Matthias Dehmer gratefully acknowledges financial support from the
German Federal Ministry of Education and Research (BMBF) (project
RiKoV, Grant No. 13N12304).

\bibliographystyle{abbrv}

\end{document}